\documentclass[%
reprint,
 amsmath,amssymb,
 aps,
prl
]{revtex4-1}

\usepackage{graphicx}
\usepackage{dcolumn}
\usepackage{bm}

\usepackage[usenames]{xcolor}

\begin{document}

\title{Nuclear and dark matter heating in massive white dwarf stars}

\author{C. J. Horowitz}
\email{horowit@indiana.edu}
\affiliation{Center for Exploration of Energy and Matter and Department of Physics, Indiana University, Bloomington, IN 47405, USA}

\date{\today}

\begin{abstract}
Recently, Cheng {\it et al.} identified a number of massive white dwarfs (WD) that appear to have an additional heat source providing a luminosity near $\approx 10^{-3}L_\odot$ for multiple Gyr \cite{Cheng_2019}.   In this paper we explore heating from electron capture and pycnonuclear reactions.  We also explore heating from dark matter annihilation.  WD stars appear to be too small to capture enough dark matter for this to be important.  Finally, if dark matter condenses to very high densities inside a WD this could ignite nuclear reactions.  We calculate the enhanced central density of a WD in the gravitational potential of a very dense dark matter core.  While this might start a supernova, it seems unlikely to provide modest heating for a long time.  We conclude that electron capture, pycnonuclear, and dark matter reactions are unlikely to provide significant heating in the massive WD that Cheng considers. 
\end{abstract}

\maketitle

Can low level nuclear reactions, or dark matter annihilation, heat massive white dwarf stars (WD)?   Recently Cheng {\it et al.} identified a number of WD, with masses between 1.08 and 1.23M$_\odot$, that appear to have an additional heat source.   This extra heat may maintain the star's luminosity near $\approx 10^{-3}L_\odot$ for multiple Gyr \cite{Cheng_2019}.  Latent heat from crystallization \cite{crystallization,Winget_2009,PhysRevLett.104.231101} and gravitational energy released from conventional $^{22}$Ne sedimentation \cite{Bildsten_2001,PhysRevE.82.066401} do not appear to be large enough to explain this luminosity.  Note that $^{22}$Ne sedimentation is significantly slowed down by C/O crystallization \cite{PhysRevE.86.066413,PhysRevE.84.016401} however Blouin et al. speculate that Ne phase separation could enhance the heating from conventional Ne sedimentation \cite{2020arXiv200713669B}.

In this paper, we explore heating from electron capture reactions, see for example \cite{PhysRevLett.123.262701}, and pycnonuclear (or density driven) fusion reactions \cite{1969ApJ...155..183S,PhysRevC.74.035803,pycnoHorowitz}.  We assume isolated stars that are not in binary systems.  We are interested in reactions that may take place preferentially at the very high central densities of massive WD and may be less important at lower densities in less massive stars.  In principle, even relatively slow nuclear reactions could contribute noticeable heat.  This is because, in the absence of nuclear reactions, there is only a modest luminosity from WD cooling.    
Alternatively, dark matter annihilation in massive WD could produce additional heating, see for example \cite{PhysRevD.98.115027,PhysRevD.91.103514,PhysRevD.92.063007}.   Dark matter can produce noticeable heating even when the dark matter is made of particles with properties, such as scattering cross sections and masses that may be difficult to observe in laboratory experiments.  Furthermore, massive WD have large escape velocities.  These stars may trap lower mass, higher velocity, dark matter particles that can escape less massive stars. 

Finally, dark matter could collect in massive WD.  If this dark matter concentrates to very high densities, its gravity can modify the structure of a WD and increase the star's central density.  This in turn could further increase the rate of electron capture and or pycnonuclear fusion reactions.

\begin{figure}[ht]
\smallskip
\includegraphics[width=1.0\columnwidth]{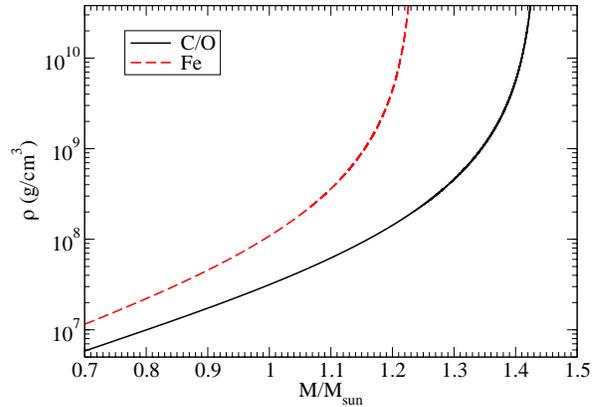}
 \caption{Central density of carbon / oxygen (solid black line) and Fe (dashed red line) white dwarfs versus mass.}
\label{Fig1}
\end{figure}

The central density $\rho_C$ of massive WD follows from hydrostatic equilibrium and an equation of state dominated by relativistic electrons.  In Fig. \ref{Fig1} we plot $\rho_C$  of a WD with electron fraction $Y_e=0.5$, this could be made of C and O or O and Ne.  We also show $\rho_C$ for a possible Fe WD with $Y_e\approx 0.464$.    We assume a simple relativistic free Fermi gas equation of state and neglect Coulomb corrections.  The central density of a C/O WD can exceed $10^9$ g/cm$^3$ for star masses above 1.35M$_\odot$.   
 
High densities can drive electron capture reactions, see for example \cite{1990A&A...227..431H}.  In Table \ref{Table1} we list the threshold densities $\rho_T$ for a variety of electron capture reactions.  This density is where the electron Fermi energy is high enough to provide for the reaction Q value.  We calculate $\rho_T$ from atomic masses.  In general, the threshold density is seen to decrease as the mass number increases.  For C/O or O/Ne stars, electron capture (at zero temperature) is not expected until $\rho_C>\rho_T\approx 6\times10^9$ g/cm$^3$ and this density is not reached until the mass of the star is above 1.40M$_\odot$, see Table \ref{Table1}.  

The threshold densities in Table \ref{Table1} are for ground state to ground state transitions.  These transitions may be forbidden by the high spin of the daughter nucleus.  However, a large forbidden matrix element was recently observed for the transition corresponding to electron capture from the ($0^+$) $^{20}$Ne ground state to the ($2^+$) $^{20}$F ground state \cite{Ne20}.  If the reaction must proceed via an excited state of the daughter nucleus, to obtain a significant rate, $\rho_T$ will be even higher than the value in Table \ref{Table1}.

\begin{table}[ht]
\begin{tabular}{cccc}
Reaction & $\rho_T$(g/cm$^3$) & $Y_e$ & $M$($M_\odot$)\\
 \hline
 ${}^{12}{\rm C}(e,\nu){}^{12}{\rm B}$ & $3.9\times 10^{10}$& 0.5 & 1.42\\
 ${}^{16}{\rm O}(e,\nu){}^{16}{\rm N}$ & $1.9\times 10^{10}$ & 0.5 & 1.42\\
 ${}^{20}{\rm Ne}(e,\nu){}^{20}{\rm F}$ & $6.2\times 10^9$ & 0.5 & 1.40 \\
${}^{56}{\rm Fe}(e,\nu){}^{56}{\rm Mn}$ & $1.1\times 10^9$ & 0.464 & 1.16 \\
${}^{54}{\rm Fe}(e,\nu){}^{54}{\rm Mn}$ & $2.1\times 10^7$ & 0.464 & 0.80\\
${}^{12}{\rm C}({}^{12}{\rm C},\alpha){}^{20}{\rm Ne}$ & $\approx 3\times 10^9$& 0.5 & 1.38 \\
 
\end{tabular}
\caption{Threshold density $\rho_T$ for the indicated nuclear reaction in a star of electron fraction $Y_e$.  The final column is the mass $M$ of a WD with central density equal to $\rho_T$. }
\label{Table1}
\end{table}

Cheng {\it et al.} considered C/O or O/Ne WD with masses between 1.08 and 1.23M$_\odot$ \cite{Cheng_2019}.  They inferred these masses by comparing the stars absolute magnitudes and colors to WD cooling models.  Note that the absolute magnitudes were determined by recent Gaia parallax measurements \cite{Gaia}.    The central density of a 1.23M$_\odot$ WD (assuming $Y_e=0.5$) is only $1.9\times 10^8$ g/cm$^3$.  This is too low for electron capture on $^{12}$C, $^{16}$O, or $^{20}$Ne, see Table \ref{Table1}.  Therefore conventional electron capture reactions are likely not significantly heating the stars Cheng considers.

It may be possible to form WD with Fe cores where the electron fraction is $Y_e\approx$26/56=0.464 \cite{FeWD}.  For example a failed SN could leave behind an Fe core \cite{FailedSN}.  Not only do these stars have higher central densities, see Fig. \ref{Fig1}, but the threshold density for electron capture on Fe is also lower $\approx 1.1\times10^9$ g/cm$^3$, see Table \ref{Table1}.  This density is reached in a 1.16M$_\odot$ Fe WD.  Furthermore, impurities could have even lower threshold densities.  For example, $^{54}$Fe has a sizable isotopic abundance on Earth $\approx 6\%$ and a very low threshold density for electron capture, see Table \ref{Table1}.  We conclude that electron capture could very well be significant in massive Fe WD.

In addition to electron capture, pycnonuclear, or density driven, fusion reactions can also take place \cite{1969ApJ...155..183S}, see for example \cite{PhysRevC.74.035803}.  In pycnonuclear fusion, quantum zero point motion allows two nuclei to approach and tunnel through the coulomb barrier.  The pycnonuclear fusion reaction that occurs first, at the lowest density, is likely to be $^{12}$C + $^{12}$C.  This is because heavier nuclei, in general, will need to tunnel through larger coulomb barriers.  Pycnonuclear reactions are greatly aided by the strong screening of the coulomb barrier by other nearby ions \cite{pycnoHorowitz}.   At present there are significant uncertainties in pycnonuclear reaction rates because they depend very sensitively on the exact distribution of ions within the crystal lattice.  In addition, there is some uncertainty in the nuclear S factor at very low energies, see for example Ref. \cite{PhysRevLett.124.192701}.

Nevertheless, there are useful estimates of pycnonuclear rates either in the pure pycnonuclear regime near zero temperature or in the thermally enhanced pycnonuclear regime at somewhat higher temperatures. In order to have a luminosity near $10^{-3}$ L$_\odot$ from $^{12}$C+$^{12}$C fusion we estimate needing a reaction rate of roughly $R\approx 5\times 10^{11}$ cm$^{-3}$s$^{-1}$.  Using the rate shown in the insert to Fig. 3 of Ref. \cite{PhysRevC.74.035803} this requires a density of very roughly $\rho_T\approx 3\times 10^9$g/cm$^3$ as listed in Table \ref{Table1}.  This density is an order of magnitude larger than the $2\times 10^8$ g/cm$^3$ central density of a 1.25M$_\odot$ (C/O) WD.  Although there is considerable uncertainty in the pycnonuclear rate, it is unlikely the uncertainty is this large.  Furthermore, the pycnonuclear rate depends strongly on the density.  Therefore, if pycnonuclear fusion were to provide $10^{-3}$L$_\odot$ for a 1.25M$_\odot$ star, the luminosity would likely be very much smaller for even slightly smaller stars and very much larger for even slightly more massive stars.  We conclude that pycnonuclear fusion is unlikely to provide significant heating for many of the massive WD that Cheng {\it et al.} considers \cite{Cheng_2019}.   

Instead of conventional nuclear reactions, the annihilation of dark mater could heat massive WD, see for example \cite{PhysRevD.98.115027,PhysRevD.91.103514,PhysRevD.92.063007}.  Furthermore, massive WD have large escape velocities and therefore they may be able to trap light dark matter particles that might escape lower mass stars, see for example \cite{Dasgupta_2019}.  However WD are small and there is likely an upper limit on the rate that they can trap dark matter particles.  An upper limit on the capture cross section $\sigma$ is provided by assuming that every particle that contacts the star is captured.  This yields,
\begin{equation}
\sigma\le \pi b_{max}^2\, .
\end{equation}
Here  $b_{max}$ is the maximum impact parameter that just collides with a star of radius $R_*$.  Conservation of angular momentum gives $b_{max}v_d=R_*(v_e^2+v_d^2)^{1/2}$ for dark matter of velocity $v_d$  and $v_e$ is the star's escape velocity.  In the solar neighborhood, we assume $v_d\approx 220$km/s.  The capture cross section is bounded by,
\begin{equation}
\sigma\le \pi R_*^2(1+v_e^2/v_d^2)\, .
\end{equation}
If the dark matter particles then annihilate, with a rate that is in steady state equilibrium with the capture rate, the luminosity from dark matter heating $L_D$ will be,
\begin{equation}   
L_D=\sigma v_d \rho_d c^2 \le \pi R_*^2 (1+v_e^2/v_d^2)v_d \rho_d c^2\, .
\label{eq.Ld}
\end{equation}
Here the dark matter density is $\rho_d\approx 0.4$ GeVc$^{-2}cm^{-3}$.   Note that Eq. \ref{eq.Ld} is independent of the mass of the dark matter particles.   For a 1.2 M$_\odot$ carbon / oxygen WD with $R_*=4\times 10^8$ cm and $v_e=9\times 10^8$ cm/s we have,
\begin{equation}
L_D\le 3\times 10^{-9}\ L_\odot\, . 
\label{eq.Ld9} 
\end{equation}
This is much less than the $10^{-3}L_\odot$ considered by Cheng et al.  We conclude that heating from dark matter annihilation is unlikely to be important in these stars.  Note that dark matter heating may be more important for WD in a globular star cluster with lower $v_d$ and possibly larger dark matter density $\rho_d$ \cite{PhysRevD.91.103514}.

White dwarfs are small and therefore may only capture dark matter at a low rate.  Perhaps more dark matter could have been captured during the main sequence phase when the star was much larger.   Let a star have a mass $M_d$ of dark matter that was previously accumulated and we now assume that the dark matter does not annihilate.  This matter may collect near the center of the star until it reaches a local dark matter density $\bar\rho_d$.  This density may depend on the nature of the dark matter such as the mass and nature of the constituent particles and the core temperature of the WD.  If $\bar\rho_d$ is comparable to or less than the central density of the star, we expect the dark matter to make almost no change in the structure of the star because we assume  $M_d\ll M_\odot$.   However if the dark matter collapses until it is very dense so that $\bar\rho_d$ becomes much larger than the original central density, gravity from the dark matter will increase the star's central density.  This change could increase the rate of electron capture or pycnonuclear fusion and may ignite a supernova, see for example \cite{PhysRevD.92.063007, PhysRevD.100.035008}.

We now calculate the density of conventional matter in a star with a dense dark matter core.  We assume the dark matter core is spherical with constant density $\bar\rho_d$, radius $r_d$ and total mass $M_d=4\pi\bar\rho_d r_d^3/3$.  For simplicity we parametrize the distribution of dark matter in terms of a constant density $\bar\rho_d$ and we try to avoid making detailed assumptions regarding the microphysics that might give rise to $\bar\rho_d$.   
  
A WD in hydrostatic equilibrium has a pressure gradient $dP/dr$ that satisfies,
\begin{equation}
\frac{dP}{dr}=-\frac{GM_{\rm tot}(r)}{r^2}\rho(r)\, .
\label{eq.dpdr}
\end{equation}
Here the density of conventional matter is $\rho(r)$ and the total enclosed mass $M_{\rm tot}$ includes dark matter contributions,
\begin{equation}
M_{\rm tot}(r)=4\pi\int_0^r r'^2dr'[\rho(r)+\bar\rho_d\Theta(r_d-r')]\, .
\end{equation}
Writing $dP/dr=d\rho/dr/(d\rho/dP)$ and integrating Eq. \ref{eq.dpdr} from the surface of the star at radius $R^*$ to the center where $\rho(0)=\rho_c$ yields,
\begin{equation}
\int_0^{\rho_c}\frac{d\rho}{\rho\frac{d\rho}{dP}} = -\int_{R^*}^{0}\frac{GM_{\rm tot}(r)}{r^2}\, dr\, .
\label{eq.drhodr}
\end{equation}
The right hand side of this Eq., after integrating by parts, can be written in terms of the gravitational potential difference between the surface and center of the star $\Delta \phi=\phi(0)-\phi(R^*)$.  If one neglects dark matter for the moment, one has,
\begin{equation}
\Delta\phi =4\pi G\int_0^{R^*} dr' r'\rho(r') -GM/R^*\, .
\end{equation}
Here $M$ is the mass of conventional matter in the star.  For the left hand side of Eq. \ref{eq.dpdr} we use the equation of state of an electron Fermi gas \cite{koonin},
\begin{equation}
\frac{d\rho}{dP}=\frac{m_0}{Y_em_e}\frac{3(1+x^2)^{1/2}}{x^2}\, .
\label{eq.eos}
\end{equation} 
Here $m_0$ is the atomic mass unit, $Y_e$ the electron fraction and $m_e$ the electron mass.  The ratio of the electron Fermi momentum to its mass is $x=(\rho/\rho_0)^{1/3}$ with $\rho_0=m_0m_e^3/(3\pi^2Y_e)$.  Using Eq. \ref{eq.eos}, the integral on the right hand side of Eq. \ref{eq.drhodr} can be evaluated to yield,
\begin{equation}
m_e[(1+(\rho_c^0/\rho_0)^{2/3})^{1/2}-1]=\frac{m_0}{Y_e}\Delta\phi\, .
\label{eq.rhocphi}
\end{equation}
Here $\rho_c^0$ is the central density of the star without any dark matter. This equation reflects energy conservation.  The electron Fermi energy at the center of the star minus the Fermi energy at the surface is equal to the mass per electron $m_0/Y_e$ times the difference in gravitational potential.  
   
We now add a small mass of dark matter $M_d$.  We do not expect the density though out most of the star to change very much because $M_d\ll M$.  Only very near the center will the gravitational potential change significantly because of the dark matter.  The $\Delta \phi$ on the right hand side of Eq. \ref{eq.rhocphi} will be increased $\Delta\phi\rightarrow\Delta\phi+\phi_{\rm dark}$.  Here the gravitational potential at $r=0$ of the dark matter core is,
\begin{equation}
 \phi_{\rm dark}(0)=\frac{3GM_d}{2r_d}\, .
 \label{eq.phidark}
 \end{equation}    
Adding this to $\Delta\phi$ in Eq. \ref{eq.rhocphi} gives the new central density (of conventional matter) $\rho_c$ in the presence of the dark matter,
\begin{equation}
m_e[(1+(\rho_c/\rho_0)^{2/3})^{1/2}-1]=\frac{m_0}{Y_e}[\Delta\phi+\phi_{\rm dark}(0)]\, .
\label{eq.rhocphidark}
\end{equation} 
Eliminating $\Delta\phi$ between this Eq. and Eq. \ref{eq.rhocphi}, assuming $\rho_c^0\gg\rho_0$, and replacing $r_d=[3M_d/(4\pi\bar\rho_d)]^{1/3}$ in favor of $\bar\rho_d$ in Eq. \ref{eq.phidark} gives,
\begin{equation}
\rho_c =\Bigl[{\rho_c^0}^{1/3}+\Bigl(\frac{3}{2\pi\, Y_e^4}\Bigr)^{1/3}Gm_0^{4/3}M_d^{2/3}\bar\rho_d^{1/3}  \Bigr]^3\, ,
\label{eq.rhoc}
\end{equation}
or,
\begin{equation}
\rho_c = \Bigl[\, {\rho_c^0}^{1/3}+1.297 \Bigl(\frac{0.5}{Y_e}\Bigr)^{4/3}\Bigl(\frac{M_d}{M_\odot}\Bigr)^{2/3}\bar\rho_d^{1/3}\, \Bigr]^3\, .
\end{equation}
This equation shows how the central density of a star with dark matter  $\rho_c$ is enhanced compared to the central density of the original star $\rho_c^0$  by the gravitational potential from the dark matter core.  We have neglected relativistic corrections to the gravitational potential.    These may only be very large as one approaches the density of a Schwartzchild black hole.  This is $1.8\times 10^{16}(M_\odot/M_D)^2 $g/cm$^3$ or $1.8\times10^{24}$ g/cm$^3$ for   $M_D=10^{-4}M_\odot$.
\begin{figure}[hbt]
\smallskip
\includegraphics[width=1.0\columnwidth]{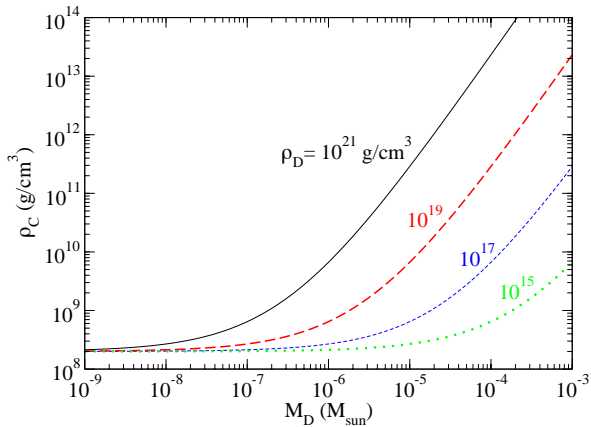}
 \caption{Central density of an originally 1.25 $M_\odot$ WD vs the mass of a dark matter core.  The curves, right to left, assume a dark matter core density $\bar\rho_d$ of $10^{15}$ to $10^{21}$ g/cm$^3$.}
\label{Fig2}
\end{figure}

Figure \ref{Fig2} shows $\rho_c$ for a 1.25 $M_\odot$ WD (with $Y_e=0.5$).  The original central density is $\rho_c^0=2\times 10^8$ g/cm$^3$.  Unless the dark matter core is very dense, the star's central density will be little changed because we assume $M_d\ll M_\odot$.  However, the central density will increase for a very dense dark matter core.   For example, pycnonuclear fusion may occur at densities of order $3\times 10^9$ g/cm$^3$ (Table \ref{Table1}).   This density will be reached for dark matter masses and densities that satisfy,
\begin{equation}
\Bigl(\frac{M_d}{10^{-6}M_\odot}\Bigr)^2\Bigl(\frac{\bar\rho_d}{10^{20} {\rm g/cm}^3}\Bigr)\ge 1\, .
\end{equation}
Note that the central density of the star does not depend on $M_d$ and $\bar\rho_d$ separately but only on the combination $M_d^2\bar\rho_d$.

\begin{figure}[ht]
\smallskip
\includegraphics[width=1.0\columnwidth]{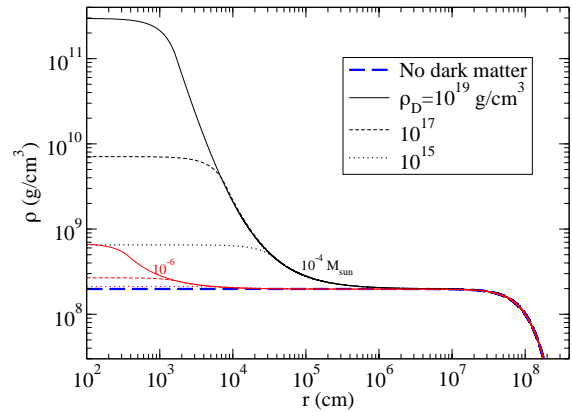}
 \caption{Density profile of an originally 1.25 $M_\odot$ WD vs radius $r$.  The dashed blue line has no dark matter while the red lines assume $M_d=10^{-6}M_\odot$ and the black line $10^{-4}M_\odot$.  The density of the dark matter core is $10^{15}$ (dotted), $10^{17}$ (dashed) and $10^{19}$ g/cm$^3$ (solid).  Note the log scales.}
\label{Fig3}
\end{figure}

Figure \ref{Fig3} shows the radial profile of a 1.25 $M_\odot$ star.  This was obtained by numerically integrating Eq. \ref{eq.dpdr}.  Although the central density only depends on the combination $M_d^2\bar\rho_d$, the size of the high density region is seen to grow with increasing $M_d$.  Given the log scale in Fig. \ref{Fig3}, we note that there is only a tiny amount of (conventional) matter in the high density central region.

In summary, if dark matter condenses to very high densities inside a WD then this will also increase the density of conventional matter and could start pycnonuclear or electron capture reactions.  What happens next may depend on the dynamical scenario.  One possibility is the ignition of a Type Ia supernova and the complete destruction of the star.  Another possibility, if the high density region is very small indeed, is that the tiny amount of material in this region is burned to Fe without releasing enough heat to start material burning at the lower densities outside the small dark matter core.  In this case the dark matter may become encased in a more or less inert Fe core with little overall change to the star.  In neither case would there be a modest amount of heat for billions of years.

In conclusion, Cheng {\it et al.} identified a number of WD, with masses between 1.08 and 1.23M$_\odot$, that appear to have an additional heat source providing a luminosity near $\approx 10^{-3}L_\odot$ for multiple Gyr \cite{Cheng_2019}.  Indeed massive WD are interesting because of their high central densities.  In this paper we explored heating from electron capture and pycnonuclear reactions.  These reactions appear to need higher densities than the central density of a $1.23M_\odot$ star.  We also explored heating from dark matter annihilation.  We found that WD appear to be too small to capture enough dark matter for this to be important.  Finally, if dark matter condenses to very high densities inside WD this could ignite nuclear reactions.  While this might start a supernova, it seems unlikely to provide modest heating for a long time.  We conclude that electron capture, pycnonuclear, and dark matter reactions   are unlikely to provide significant long term heating in massive WD.

\acknowledgements{We thank Ed Brown, Matt Caplan and Brendan Reed for very helpful comments.  This work is supported in part by United States Department of Energy Office of Science, Office of Nuclear Physics grants DE-FG02-87ER40365 and DE-SC0018083.} 


\end{document}